# Coplanar Lateral Gating $MoS_2$ on $SrTiO_3$: A Unified Platform for Classical and Quantum Devices

*Prasad Muragesh[#], Manav Murali[#], Venkatesha Modur Ramachandra, and Madhu Thalakulam[*]*

Indian Institute of Science Education and Research Thiruvananthapuram, Thiruvananthapuram, Kerala 695551, India

## ABSTRACT

Atomically thin transition-metal dichalcogenides, such as $MoS_2$, offer a promising route to surpass the scaling limits of silicon owing to their excellent electrostatic control and resilience to short-channel effects. Realizing this potential depends critically on gate-stack engineering, where strong gate coupling must be achieved without compromising the pristine two-dimensional interface. Here, we demonstrate a coplanar lateral-gating architecture for $MoS_2$ field-effect transistors fabricated directly on single-crystal $SrTiO_3$. The exceptionally high permittivity of $SrTiO_3$ enhances gate–channel coupling. The coplanar geometry eliminates the need for a separate gate insulator, reducing interface disorder. Owing to the quantum paraelectric nature, the $SrTiO_3$ dielectric constant increases upon cooling, enhancing the gate coupling, leading to a decrease in threshold voltage and subthreshold swing an effect that contrasts with conventional FET architectures. The dielectric-free $MoS_2$ surface, combined with enhanced electrostatic control, makes this architecture a promising platform for low-power two-dimensional electronics and cryogenic quantum devices.

[#] equally contributed to the work

[*] madhu@iisertvm.ac.in

Gating technology and architecture are central to both classical and quantum semiconductor technologies because they provide the electrostatic control required to modulate carrier density, conductivity, and overall device performance. The evolution from planar MOSFETs to FinFETs and gate-all-around architectures reflects this continuing pursuit of stronger gate-to-channel electrostatic control[1,2]. As device dimensions continue to shrink, the ability of the gate to control an increasingly confined channel becomes a non-trivial factor governing device performance, scaling, short-channel effects, leakage current, and power consumption. In this context, the scaling length of a semiconductor device suggests the need for very thin gate dielectrics and semiconducting channel, with a high dielectric constant for the gate insulator and a comparatively low dielectric constant for the semiconductor. Owing to their atomically thin crystalline defect-free channel, $MoS_2$ and other semiconducting transition-metal dichalcogenides (TMDCs), offer a promising route beyond the scaling limits of conventional silicon-based CMOS technology. Monolayer metal-oxide-semiconductor field-effect transistors (MOSFETs)[3], logic circuits[4,5], and high-frequency transistors[6], have been demonstrated on $MoS_2$. Intel recently developed a 300 mm FAB process for $MoS_2$ , $WS_2$, $WSe_2$, and $MoSe_2$ at CMOS-compatible temperatures[7], demonstrating complementary $MoS_2$ n-FET and $WSe_2$ p-FET integration. TSMC reported nanoscale $WS_2$ p-FETs using area-selective CVD, while IMEC demonstrated functional $WS_2$ FETs on 300 mm Si substrates with FAB-compatible flows[8]. In addition, demonstrations of important quantum transport phenomena such as Shubnikov-de Haas oscillations[9,10], Coulomb blockade[11–13] , and conductance quantization[14–17], electron-spin-resonance[18,19], Moire-minibands[20] also make $MoS_2$ a promising candidate for quantum technological applications. Recent demonstrations of gate-defined quantum dots and spin-valley states have established key building blocks for future qubit implementations, positioning $MoS_2$ as a promising material platform for both beyond-CMOS electronics and emerging quantum technologies.

However, realizing this potential critically depends on the gate architecture and the properties of the gate insulator, particularly its dielectric constant, physical thickness, and semiconductor–dielectric interface quality. The gate capacitance which is directly proportional to the permittivity of the gate-dielectric, directly determines the efficiency with which an applied gate voltage modulates carrier density and channel potential. Mono or few-layered $MoS_2$ channel with a high-$\kappa$ gate-dielectric therefore offer a compelling route to device scaling[21]. Unlike bulk semiconductors, where charge transport occurs predominantly away from the surface, the entire conducting channel in atomically thin TMDCs lies in direct proximity to the supporting substrate and gate dielectric, making interface quality one of the most important parameters. Historically, most $MoS_2$ transistors have been fabricated on thermally grown $SiO_2$ on highly doped silicon[22] because this platform is low-cost, CMOS compatible, and enables global electrostatic gating through a doped silicon back gate. For local gate-control, high-κ dielectrics such as $Al_2O_3$ and $HfO_2$ are commonly used[23], but these materials possess high densities of charge traps, adsorbates, and remote phonon scattering sources that reduce carrier mobility, introduce charge noise, and limit quantum coherence. In addition, metal-induced gap states, Fermi-level pinning, and interfacial contamination often produce Schottky barriers and elevated contact resistance, thereby limiting carrier injection, suppressing mobility, and obscuring the intrinsic transport properties of the channel[24,25]. These challenges are even more significant in quantum devices operating at cryogenic temperatures, where charge fluctuations and dielectric disorder directly affect the device stability, coherence times, and qubit fidelity[26,27]. Consequently, creating high-quality dielectric/TMDC interfaces and gate-stack engineering are essential for enabling technology for future high-performance, low-power, and quantum-enabled $MoS_2$/TMDC semiconductor devices.

Here, we present a strategy for achieving strong gate control across a broad temperature range—from room temperature down to the cryogenic regime—without compromising channel quality or interface integrity. We use strontium titanate ($SrTiO_3$ or STO) as the dielectric substrate together with a lateral-gating architecture, a configuration commonly employed in quantum devices such as gate-defined quantum dots and quantum point contacts[28–31]. STO has a high dielectric constant of approximately 300 at room temperature, which increases to about 25,000–30,000 below 4 K[32–35]. This exceptional permittivity confines the electric field effectively within the substrate, making lateral gating a particularly efficient strategy. Crucially, because the substrate itself is crystalline and no top-gate dielectric is required, the $MoS_2$ interface remains pristine and defect-free. This architecture delivers excellent electrostatic control over the channel while preserving high carrier mobility across the entire temperature range studied. We further observe a marked reduction in subthreshold swing (SS) at cryogenic temperatures—a direct consequence of STO's rising dielectric constant as temperature decreases. This inverse relationship between temperature and gating efficiency is especially advantageous: as devices are cooled toward the quantum regime, gate control only improves. Together, these results establish lateral-gated STO as a compelling and scalable platform for both classical and quantum device architectures at cryogenic temperatures.

Fig 1. (a) is the schematic representation of the device. The devices are fabricated on 500 $\mu m$ thick commercially available (100)-oriented STO substrates. $MoS_2$ flakes were prepared by micromechanical exfoliation from $MoS_2$ single crystals[36]. The flakes are then transferred using PDMS film by deterministic in-house developed transfer setup[37]. A layer of hexamethyldisilazane (HMDS) was spin coated on the standard cleaned STO substrates, prior to $MoS_2$ transfer to passivate the surface for better adhesion and to minimize the interface trap effects[38]. The device architecture consists of two types of gates, lateral or in-plane and the conventional back-gate. The lateral gates and the source/drain contacts are defined by electron-beam lithography followed by Cr/Au metallization. Back gate (BG) is facilitated using a uniform coating of silver epoxy on the backplane of the STO crystal. The lateral gates consist of two types of gates: the side gates (SG) and contact gates (CG). Contact-gates (CG1, CG2, CG3 and CG4) are designed to predominantly

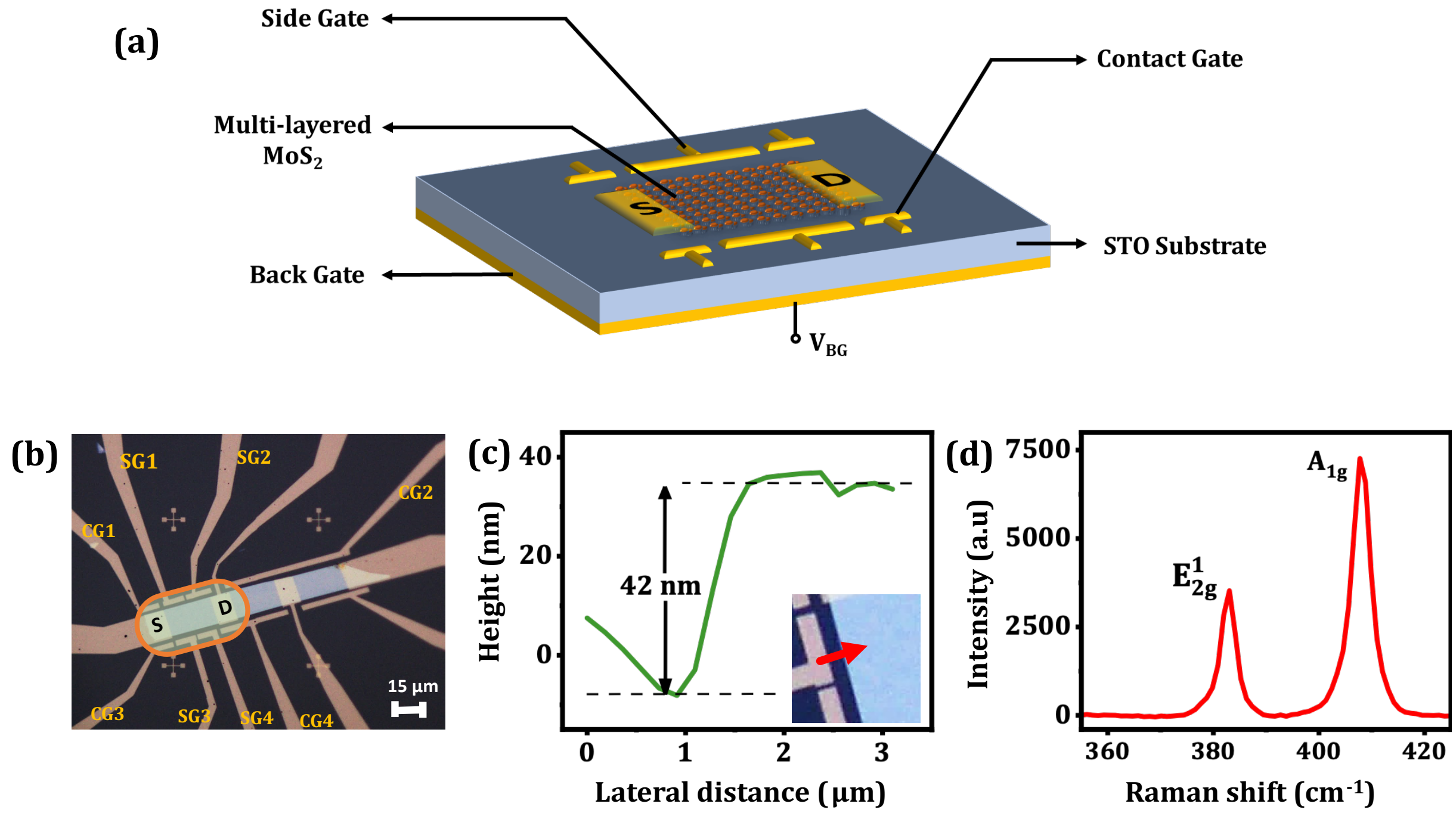

**Figure 1**. **Device architecture.** (**a**) Schematic representation of the lateral-gate architecture for $MoS_2$ on STO. (**b**) Optical image of the fabricated device on STO; the highlighted region indicates the portion of the device used for the measurements. (**c**) Atomic force microscope (AFM) height profile at the edge of the $MoS_2$ flake and thickness of about 42 $nm$ corresponding to a multilayer flake ~60 layers. (**d**) Raman spectrum of exfoliated multilayer $MoS_2$ on STO, the two characteristic peaks of $MoS_2$, $E_{2g}^1$ and $A_{1g}$, are separated by 24.72 $cm^{-1}$ which corresponds to multilayer $MoS_2$.

dope the $MoS_2$ region under the source/drain contacts and reduce the Schottky barrier, while side-gates (SG1, SG2, SG3 and SG4) are employed to electrostatically dope the channel region and control the channel conductance. On the device discussed in the main article, all lateral gates are positioned ~ $1.5\ \mu m$ away from the $MoS_2$ channel. We have studied a total of three devices with similar architecture (see Supporting Information, SI-1); the discussion in the main manuscript is focussed on the device shown in Fig. 1(b). All the data presented in this manuscript is obtained from the highlighted region in Fig. 1(b).

Fig. 1(c) shows the atomic force microscope height profile at the edge of the $MoS_2$ flake (see the inset) indicating a thickness of about $42\ nm$ corresponding to a multilayer flake (~60 layers)[39]. Fig. 1(d) shows the Raman spectrum of the $MoS_2$ flake presented in Fig. 1(b). The Raman spectrum shows the signature $E_{2g}^{1}$ and $A_{1g}$ lines at $383.04\ cm^{-1}$ and $407.76\ cm^{-1}$ respectively, with a peak separation of $24.72\ cm^{-1}$, corresponding to a multilayer $MoS_2$ sample[40], implying that the STO substrate has not modified the flake characteritics. All electrical measurements are performed in high-vacuum, inside a closedcycle cryostat, in the temperature range between room-temperature and 10 K.

Fig. 2 (a) shows the effect of the back-gate voltage $V_{BG}$ on the current-voltage (I-V) characteristics of the device; $V_{BG}$ is varied between $-30\ V$ to $+30\ V$ while all the lateral gates are maintained at 0 V. The Ohmic nature of the source and drain contact is evident from the linear behaviour of the traces. The I-V characteristics clearly show an increase in the device conductance with $V_{BG}$ suggesting n-type nature of the $MoS_2$ flake[41,42]. The top-inset to Fig. 2(a) presents the transfer characteristics with respect to $V_{BG}$ for a source-drain voltage $V_{SD} = 10\ mV$. Even though the back-gate is located $\sim 500\ \mu m$ below the sample, we find a significant modulation of the source-drain current, compared to that we observe on a conventional 300 nm $SiO_2$ on highly doped silicon backgated samples [see Supporting Information, SI-2 fig (a) and (c)].

Now we inspect the effect of lateral gates on the channel conductance. Fig.2 (b) shows the I-V characteristics of the device for voltages between $-1.5\ V$ to $1.5\ V$, applied together to all the lateral gates while $V_{BG} = 0\ V$. The source-drain current shows significant increase with the voltage applied onto the lateral gates. To differentiate between the contributions of contact and side gates, we further investigate the I-V characteristics of the device as a function of contact-gate voltages while keeping $V_{SG} = 2.5$ V, as shown in inset to Fig. 2 (b). We observe all I-V traces maintain the linear behavior. Nevertheless, the increase in the source-drain current against the contact gate voltage is comparatively much smaller, suggesting that the dominant role is played by the side gates in improving the channel conductivity, observed in Fig. 2 (b).

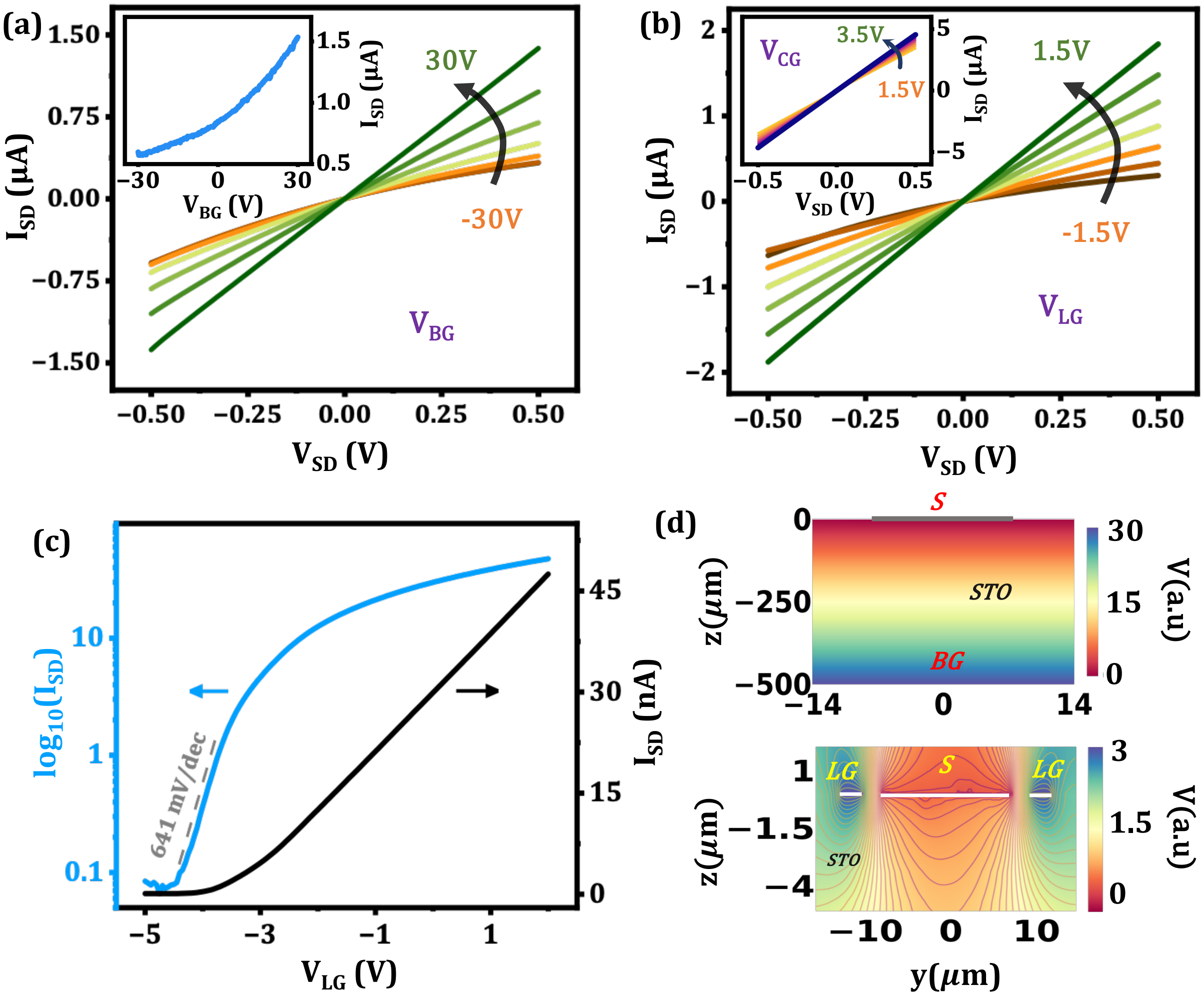

**Figure 2. Room-temperature electrical characterization**. (**a**) I-V characteristics measured for back-gate voltages ($V_{BG}$) from $-30\ V$ to $30\ V$. *Inset* - transfer characteristics with respect to the back-gate voltage. (**b**) I-V characteristics for various lateral-gate (LG) voltages. *Inset* – I-V characteristics for various contact gate (CG) voltages for side-gate voltage $V_{SG} = 2.5\ V$. (**c**) Transfer characteristics with resoect to the lateral gate (LG) voltage for $V_{SD} = 10\ mV$ in both linear (right) and logarithmic (left) scales showing an extracted subthreshold swing of $\sim 641\ mV/dec$, and threshold voltage $\sim -3.39$. **(d)** Simulated electrostatic potential map of the device with BG (top) and LG (bottom) configuration at room temperature.

Fig. 2 (c) represents the transfer characteristics obtained with respect to the voltage on all lateral gates, $V_{LG}$, keeping $V_{SD} = 10\ mV$. Consistent with the linear I-V characteristics shown in Fig. 2 (b), the transfer characteristics obtained with the lateral-gates exhibit superior current modulation compared to that observed with the back-gate, shown in the inset to Fig. 2 (a). The subthreshold swing ($SS$), the voltage required to obtain a ten-fold increase in current below the threshold voltage, described by $SS = \frac{dV_G}{dlog(I_{SD})}$, quantifies the switching efficiency of an FET[43]. An $SS$ of 641 $mV/dec$ extracted from the semi-log plot (blue color), shown in Fig 2(c), is superior to $MoS_2$ FETs[44–46] reported in the literature. We also extract a threshold voltage $V_{th} \sim -3.39\ V$ using the extrapolation in the linear regime (ELR) method [47,48].

The field effect mobility $\mu_{FE} = \frac{L}{WC_gV_D}(\frac{dI_D}{dV_G})$ is an important figure of merit characterizing the efficiency of the charge dynamics[49,50]. $L = 25.5\ \mu m$ is the channel length, $W = 15.36\ \mu m$ is the channel width, $V_{SD}$ is the source drain bias, $C_g$ is the gate capacitance and $V_G$ is the gate voltage. The $C_g$ is modelled and calculated using coplanar parallel strips for which the capacitance per unit length $C' = \varepsilon_0\varepsilon_r \frac{K(k')}{K(k)}$, where $\varepsilon_0$ and $\varepsilon_r$ is the permittivity of free space and relative permittivity of the substrate, $K = K(k)$ is the complete elliptical integral of the 1st kind [51,52]. From the data shown in Fig. 2 (c) we obtain a room temperature mobility ~ $50\ cm^2/Vs$ for the device, which is by far enhanced compared to previous reports on $SiO_2$[53–55].

From Fig 2 (a) and (b) we infer that, a lateral gate voltage of only $1.5\ V$ results in a channel current modulation of $\sim 2\ \mu A$, whereas similar enhancement in the channel current requires application of $V_{BG}$ in excess of $40\ V$. This is attributed to the enhanced electric field confinement and the resulting strong electrostatic gate control of the $MoS_2$ due to the high dielectric environment created by the STO substrate. To get further insight into this aspect, we simulate the potential profile on the sample due to both the back and the lateral gating configuration. Details of the simulation can be found in the Supporting Information, SI-3. Fig. 2 (d) top and bottom panels show the simulated potential profile in the YZ plane, due to the back-gate and lateral gate, respectively. The gate voltage assumed in the simulation for the lateral gate is $3\ V$ while that for the back-gate is $30\ V$. Despite the ten-fold applied voltage, the potential induced on the $MoS_2$ channel from the back gate is significantly weaker ($\sim -0.031\ V$) compared to that produced by the lateral-gate ($\sim 0.056\ V$). The resulting potential map clearly illustrates that the penetration of the electric field from the lateral gates into the channel region is stronger than that due to the back-gate, verifying the enhanced electrostatic coupling and effective gate control observed in Fig. 2 (b). Supporting Information, SI-2 summarises the room temperature transport studies on a similar laterally gated $MoS_2$ sample on a highly doped Silicon wafer hosting a $300\ nm$ $SiO_2$ layer. Unlike for the $MoS_2$ samples on the STO substrate, for the $SiO_2$/Si substrate, the lateral gate shows little effect on the channel conductance. In addition, the performance of the back-gate is also inferiour compared to both the back-gate and lateral gates on the STO substrate. The large dielectric constant, ~ 300, at room temperature effectively confines the field-lines into the STO. The enhanced performance of the lateral gates over the back-gate is due to its close proximity, 1.5 $\mu m$, as opposed to the back-gate which is $500\ \mu m$ below the sample. This also suggest that the lateral gating scheme proposed here can be used to sculpture the potential landscape on the sample by suitably designing the gate geometry and separation to suit a variety of applications.

In Fig. 3 we summarize the performance of lateral-gates for various temperatures between room temperature and 20 K. For all the measurements we maintained $V_{BG} = 0\ V$. Fig. 3 (a) is the I-V characteristics of the device from 300 K down to 20 K with $2.5\ V$ applied to all the lateral gates. The device shows linear I-V characteristics at higher temperatures and turns non-linear for temperatures below 60 K. $MoS_2$ being a 2D semiconductor, depletion of charge carriers and an increase in its resistance as the temperature is lowered is expected[49]. The non-linearity is attributed to the Schottky behavior of the Cr/Au source/drain contacts.

Fig. 3 (b) shows the transfer characteristics with respect to the lateral gate voltage for various temperatures between 300 K and 20 K with a $V_{SD} = 10\,mV$. The traces between 300 K and 80 K is offset by 500 $pA$ along the Y-axis for clarity. We observe an increase in the threshold voltage, $V_{th}$, as the temperature is reduced from 300 K till 70 K, represented by a progressive shift in the transfer characteristics towards the right, in Fig. 3 (b). This behavior is expected for a semiconductor; higher gate voltages are required to electrostatically dope the channel as the carriers are depleted due to a reduction in the temperature[56,57]. For temperatures below 70 K, we observe that the transfer characteristic traces are moving towards the left, a manifestation of a reduction in the $V_{th}$; similar behavior is also observed in the back gated $MoS_2$/STO, shown in Supporting Information, SI-4. This turnaround behavior in the $V_{th}$ is not an expected behavior of $MoS_2$ or other similar semiconductors on conventional substrates such as $SiO_2$/Si. $V_{th}$ represents the minimum gate voltage required to activate conduction in the channel, which is a measure of the electrostatic doping required to create enough mobile charge carriers. This is also a measure of the gate-induced channel potential which is a strong function of the dielectric constant and the gate-channel capacitance. Unlike that of conventional gate insulators such as $SiO_2$, the dielectric constant of STO, is a strong function of temperature. Between 300 K and ~ 110 K, the dielectric constant of STO remains relatively unaffected. Upon further reduction in temperature, the STO undergoes a structural phase transition, beyond which the dielectric constant steadily increases[58–60]. This results in an enhancement of the channel gating efficiency and eventually causing a turnaround in the $V_{th}$ around ~ 70 K. Overall observed behavior matches excellently with the above-mentioned picture. The precise turnaround temperate depends on the carrier depletion rate and the rate of increase of the dielectric constant of STO, both which depends on the details of the $MoS_2$ flake and the STO substrate.

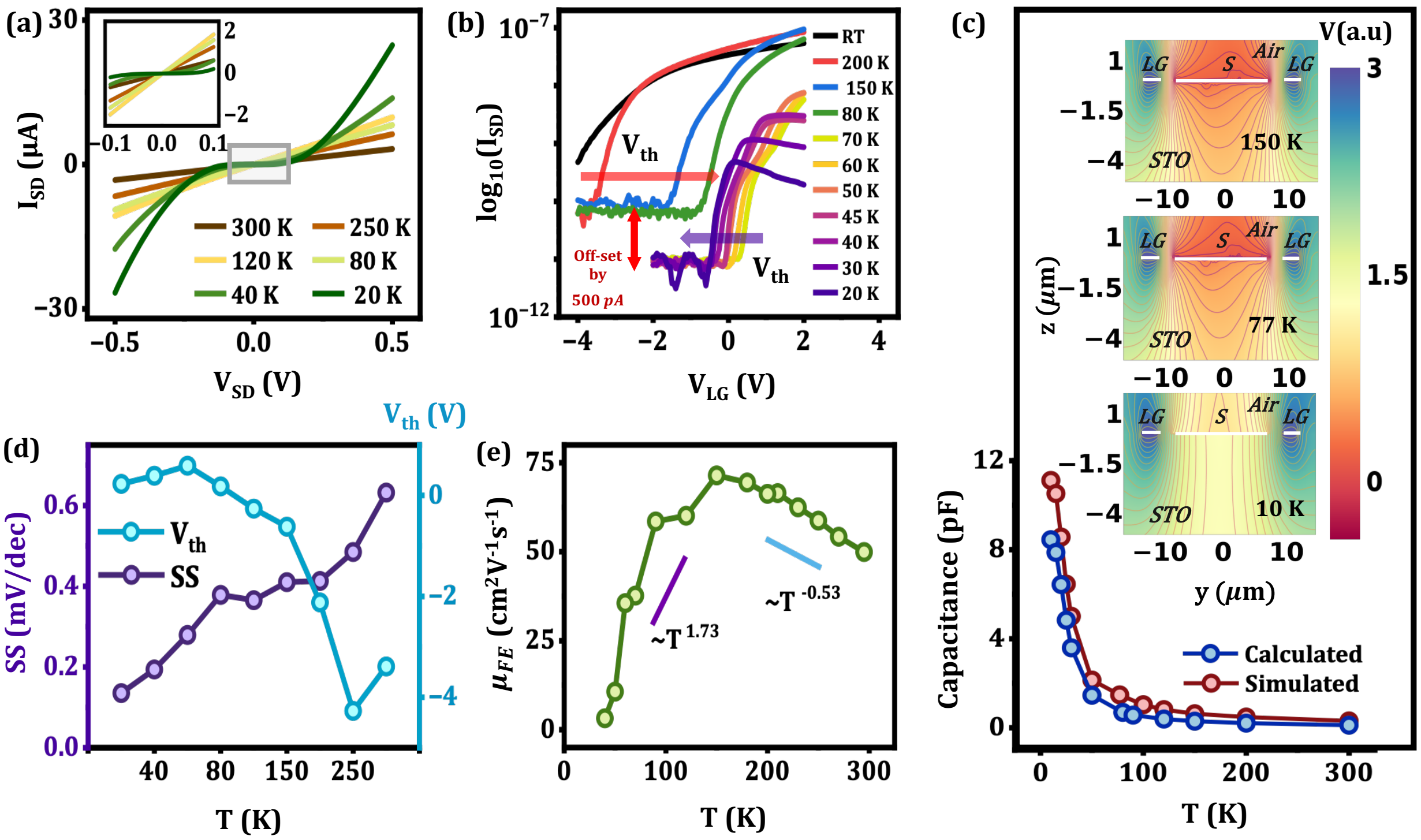

**Figure 3. Temperature-dependence and threshold-voltage reversal** (**a**) I-V characteristics recorded at $V_{SG} = 2.5\,V$ for temperatures ranging from room temperature to $20\,K$. The device gradually transitions from linear to nonlinear behaviour with decreasing temperature, primarily due to the increasingly insulating nature of $MoS_2$ at low temperatures. (**b**) Transfer characteristics woithrespect to lateral gates (LG), measured for $V_{SD} = 10\,mV$. The red and violet arrow represent the progressive shoft in the threshold voltage as the sample is cooled. (**c**) Gate-channel capacitance extracted from electrostatic simulations (red) and theoretical calculations (blue). *Inset* - electrostatic simulations showing the potential distrubutions due to the lateral gates for varoious temperatures; 150 K (top), 77K (middle) and 10 K (bottom). (**d**) Temperature dependence of the subthreshold swing and threshold voltage extracted from the transfer characteristics shown in Fig 3(b). (**e**) Field effect mobility as a function of temperature, extracted from the transfer curves in Fig 3(b), along with fitting using the power-law relation $\mu \propto T^{-n}$.

In Fig. 3(c) we verify this picture with the help of numerical simulations [see Supporting Information, SI-3 for details]. The red-trace in the main panel shows a plot of the gate-channel capacitance extracted from the simulation , which shows a slow, but gradual rise down to a temperature range ~ 110 K. For temperatures below 110 K the capacitance increases rapidly owing to the increase in STO's dielectric constant. The capacitance calculated using the coplanar parallel strip geometry is shown in blue. Both the calculated and simulated capacitance values show excellent agreement. These results are also in agreement with the experimental observations shown in Figure 3(b), providing strong evidence that the enhanced gate modulation and the shift in threshold voltage at low temperatures originate from the temperature-dependent increase in the dielectric constant of STO and the enhancement in the electrostatic doping of the $MoS_2$ channel. The insets in Fig. 3 (c) show the simulated electrostatic potential maps for $V_{LG} = 3\ V$ for three representative temperatures: 150 K (top, above the structural phase transition), 77 K (middle, slightly below the structural phase transition), and 10 K (bottom, below the transition). Simulations for additional temperatures are provided in the Supporting Information, SI-5. As the temperature decreases, the dielectric constant of STO increases significantly, particularly below the Curie temperature $T_c \sim 30$. This behaviour increases the electric-field confinement between the lateral gates and the $MoS_2$ channel and enhances the electrostatic control exerted by the lateral gates at cryogenic temperatures.

The temperature dependent $V_{th}$ extracted using ELR method from the transfer curves in Fig. 3(b) are shown in Fig 3(d) (blue, right) while the $SS$ is shown in violet (left). We find that the SS exhibits a steady decrease from $641\, mV/dec$ at 300 K to $135.3\, mV/dec$ at $20\, K$. In conventional FET technologies, minimizing the $SS$ is a key objective for achieving improved switching performance, often realized through the use of ultrahigh-κ dielectrics and negative-capacitance approaches[43,61,62]. In our case, however, the observed temperature-dependent improvement in $SS$ arises intrinsically with cooling, highlighting the potential of the device architecture for cryogenic FET applications.

Fig. 3 (e) shows the field effect mobility, for various temperatures, extracted from Fig 3(b). The mobility gradually increases from $\sim 50\, cm^2/Vs$, as the temperature is lowered attains a maximum of $71.44\, cm^2/Vs$ at 150 K. The mobility in this regime was fitted using the power-law relation, $\mu \propto T^{-n}$, where $n$ is the damping exponent that characterizes the strength of dominant scattering mechanisms [63,64]. In this regime, the damping exponent was found to be $0.53$, which is lower than the commonly expected value of 1.5 for electron-phonon scattering dominant transport[65,66]. These variation can be due to stray contributions such as homopolar phonon scattering[63]. For temperatures below $150\, K$, the mobility is found to decrease with temperature eventually reaching a value of $3.19\, cm^2/Vs$ at $40\, K$. In this regime, the damping exponent is found to be $-1.73$, close to -1.5, observed for impurity scattering limited mobility. The transport in this regime is mainly dominated by scattering from charged impurities present due to defects in the $MoS_2$ channel[67]. Additionally, this behavior of mobility against temperature is also indicative of the presence of variable range hoping (VRH)[68]. However, in multi-layered $MoS_2$, the bulk contribution must be dominant and the 2D hopping model is expected to breakdown[69]. Hence, in our device, mobility is expected to be dominated by phonons and long-range Coulomb impurities.

In conclusion, we have demonstrated that a planar, lateral-gating architecture built directly on single-crystal $SrTiO_3$ substrates offers a potential route to high-performance, interface-preserving $MoS_2$ field-effect transistors that operate reliably from room temperature down to the cryogenic regime. By exploiting STO's exceptionally large and strongly temperature-dependent dielectric constant, the lateral gates achieve electrostatic control far beyond the performances of conventional $Si/SiO_2$ back gate and high-K dielectric top-gates. This efficient coupling translates directly into device metrics that surpass most reported $MoS_2$ FETs on $SiO_2$ or STO, including a room-temperature subthreshold swing of $641\ mV/dec$ and a field-effect mobility approaching $50\ cm^2/V\cdot s$. As temperature decreases, the dielectric constant of STO rises sharply strengthening the lateral-gate coupling; this manifests as a pronounced turnaround in threshold voltage near 70 K and a continuous improvement in subthreshold swing. This behaviour — gate efficiency improving rather than degrading with cooling — runs counter to the trend seen in conventional oxide-gated FETs and is a direct, reproducible fingerprint of the STO dielectric anomaly, independently validated through temperature-dependent capacitance simulations. Taken together, these results show that lateral gating on STO decouples two properties that are normally in tension in 2D FET design: interface cleanliness and gate efficiency. The architecture requires no exotic dielectric deposition, no van der Waals stacking, and no top-gate integration, yet delivers electrostatic control and switching performance that rival or exceed devices built with ultrahigh-κ or ferroelectric top gates. Because lateral gating is also the geometry of choice for gate-defined quantum dots and quantum point contacts, the same architecture demonstrated here for classical FET metrics is directly extensible to quantum device applications, where a defect-free channel and stable, strong electrostatic confinement at cryogenic temperatures are essential. We therefore propose planar lateral gating on STO as a broadly applicable, fabrication-friendly platform for both low-power classical electronics and emerging cryogenic and quantum $MoS_2$-based technologies, readily extensible to other 2D semiconductors and to the realization of gate-defined quantum confinement on STO for future work.

**ASSOCIATED CONTENT**

**Data Availability Statement:**

The data that support the findings of this study are available within the article and its supporting information.

**Supporting Information**

Measurements and corroborating results across multiple devices, along with extended details and simulation results, are provided.

**Author Contributions:** MT conceived the problem. PM and MM prepared the devices and performed the experiments. VMR performed the simulation, PM, MM and MT co-wrote the manuscript.

**Acknowledgement:** We acknowledge National Supercomputing Mission (NSM) for providing computing resources of 'PARAM RUDRA' at IIT Patna, which is implemented by [C-DAC] and supported by MeitY and DST, Government of India. P.M. acknowledges UGC for fellowship.

**References:**

(1) Hsu, W.; Mantey, J.; Register, L. F.; Banerjee, S. K. On the Electrostatic Control of Gate-Normal-Tunneling Field-Effect Transistors. *IEEE Trans. Electron Devices* **2015**, *62* (7), 2292–2299. https://doi.org/10.1109/TED.2015.2434615.

(2) Datta, S. S.; Strachan, D. R.; Johnson, A. T. C. Gate Coupling to Nanoscale Electronics. *Phys. Rev. B* **2009**, *79* (20), 205404. https://doi.org/10.1103/PhysRevB.79.205404.

(3) Li, Y.; Qi, C.; Zhou, X.; Xu, L.; Li, Q.; Liu, S.; Yang, C.; Liu, S.; Xu, L.; Dong, J.; Fang, S.; Yang, Z.; Chen, Y.; Sun, X.; Lu, J. Monolayer ${\mathrm{WSi}}_{2}{\mathrm{N}}_{4}$: A Promising Channel Material for Sub-5-Nm-Gate Homogeneous CMOS Devices. *Phys. Rev. Appl.* **2023**, *20* (6), 064044. https://doi.org/10.1103/PhysRevApplied.20.064044.

(4) Liu, C.; Chen, H.; Hou, X.; Zhang, H.; Han, J.; Jiang, Y.-G.; Zeng, X.; Zhang, D. W.; Zhou, P. Small Footprint Transistor Architecture for Photoswitching Logic and in Situ Memory. *Nat. Nanotechnol.* **2019**, *14* (7), 662–667. https://doi.org/10.1038/s41565-019-0462-6.

(5) Zhang, X.; Zhao, H.; Wei, X.; Zhang, Y.; Zhang, Z.; Zhang, Y. Two-Dimensional Transition Metal Dichalcogenides for Post-Silicon Electronics. *Natl. Sci. Open* **2023**, *2* (4), 20230015. https://doi.org/10.1360/nso/20230015.

(6) Gao, Q.; Zhang, C.; Yang, K.; Pan, X.; Zhang, Z.; Yang, J.; Yi, Z.; Chi, F.; Liu, L. High-Performance CVD Bilayer MoS2 Radio Frequency Transistors and Gigahertz Mixers for Flexible Nanoelectronics. *Micromachines* **2021**, *12* (4), 451. https://doi.org/10.3390/mi12040451.

(7) Dorow, C. J.; Schram, T.; Smets, Q.; O’Brien, K. P.; Maxey, K.; Lin, C.-C.; Panarella, L.; Kaczer, B.; Arefin, N.; Roy, A.; Jordan, R.; Oni, A.; Penumatcha, A.; Naylor, C. H.; Kavrik, M.; Cott, D.; Graven, B.; Afanasiev, V.; Morin, P.; Asselberghs, I.; Lockhart de La Rosa, C. J.; Sankar Kar, G.; Metz, M.; Avci, U. Exploring Manufacturability of Novel 2D Channel Materials: 300 Mm Wafer-Scale 2D NMOS & PMOS Using MoS2, WS2, & WSe2. In *2023 International Electron Devices Meeting (IEDM)*; 2023; pp 1–4. https://doi.org/10.1109/IEDM45741.2023.10413874.

(8) Schram, T.; Smets, Q.; Opdebeeck, A.; Ghosh, S.; Groven, B.; Kumar, P.; Medina, H. M.; Cott, D.; deMarneffe, J.-F.; Dongre, H.; Kruv, A.; Panarella, L.; Pinotti, L. F.; Morin, P.; Kar, G. S.; Lockhart de la Rosa, C. J. Integration and Electrical Evaluation of WS2 and MoS2 Fets in a 300 Mm Pilot Line. *Discov. Electron.* **2026**, *3* (1), 15. https://doi.org/10.1007/s44291-026-00164-4.

(9) Cao, H.; Tian, J.; Miotkowski, I.; Shen, T.; Hu, J.; Qiao, S.; Chen, Y. P. Quantized Hall Effect and Shubnikov--de Haas Oscillations in Highly Doped ${\mathrm{Bi}}_{2}{\mathrm{Se}}_{3}$: Evidence for Layered Transport of Bulk Carriers. *Phys. Rev. Lett.* **2012**, *108* (21), 216803. https://doi.org/10.1103/PhysRevLett.108.216803.
(10) *Shubnikov-De Haas Effect - an overview | ScienceDirect Topics*. https://www.sciencedirect.com/topics/chemistry/shubnikov-de-haas-effect (accessed 2026-08-05).
(11) Kormányos, A.; Zólyomi, V.; Drummond, N. D.; Burkard, G. Spin-Orbit Coupling, Quantum Dots, and Qubits in Monolayer Transition Metal Dichalcogenides. *Phys. Rev. X* **2014**, *4* (1), 011034. https://doi.org/10.1103/PhysRevX.4.011034.
(12) Pisoni, R.; Lei, Z.; Back, P.; Eich, M.; Overweg, H.; Lee, Y.; Watanabe, K.; Taniguchi, T.; Ihn, T.; Ensslin, K. Gate-Tunable Quantum Dot in a High Quality Single Layer MoS2 van Der Waals Heterostructure. *Appl. Phys. Lett.* **2018**, *112* (12), 123101. https://doi.org/10.1063/1.5021113.
(13) Kotekar-Patil, D.; Deng, J.; Wong, S. L.; Goh, K. E. J. Coulomb Blockade in Etched Single- and Few-Layer MoS2 Nanoribbons. *ACS Appl. Electron. Mater.* **2019**, *1* (11), 2202–2207. https://doi.org/10.1021/acsaelm.9b00390.
(14) Phan, N. A. N.; Uddin, I.; Nazarian-Firouzabadi, A.; Chuang, C.; Chen, D.-R.; Watanabe, K.; Taniguchi, T.; Kim, G.-H. Electrostatically Confined Charge Transport in Split-Gated MoS2 Device. *Appl. Phys. Lett.* **2026**, *129* (5), 052103. https://doi.org/10.1063/5.0341242.
(15) Gold, C.; Knothe, A.; Kurzmann, A.; Garcia-Ruiz, A.; Watanabe, K.; Taniguchi, T.; Fal'ko, V.; Ensslin, K.; Ihn, T. Coherent Jetting from a Gate-Defined Channel in Bilayer Graphene. *Phys. Rev. Lett.* **2021**, *127* (4), 046801. https://doi.org/10.1103/PhysRevLett.127.046801.
(16) Sakanashi, K.; Krüger, P.; Watanabe, K.; Taniguchi, T.; Kim, G.-H.; Ferry, D. K.; Bird, J. P.; Aoki, N. Signature of Spin-Resolved Quantum Point Contact in p-Type Trilayer WSe2 van Der Waals Heterostructure. *Nano Lett.* **2021**, *21* (18), 7534–7541. https://doi.org/10.1021/acs.nanolett.1c01828.
(17) Boddison-Chouinard, J.; Bogan, A.; Barrios, P.; Lapointe, J.; Watanabe, K.; Taniguchi, T.; Pawłowski, J.; Miravet, D.; Bieniek, M.; Hawrylak, P.; Luican-Mayer, A.; Gaudreau, L. Anomalous Conductance Quantization of a One-Dimensional Channel in Monolayer WSe2. *Npj 2D Mater. Appl.* **2023**, *7* (1), 50. https://doi.org/10.1038/s41699-023-00407-y.
(18) Sharma, C. H.; Zhao, P.; Tiemann, L.; Prada, M.; Pandey, A. D.; Stierle, A.; Blick, R. H. Electron Spin Resonance in a Proximity-Coupled MoS2/Graphene van Der Waals Heterostructure. *AIP Adv.* **2022**, *12* (3), 035111. https://doi.org/10.1063/5.0077077.
(19) Sharma, C. H.; Parvangada, A.; Tiemann, L.; Rossnagel, K.; Martin, J.; Blick, R. H. Resistively Detected Electron Spin Resonance and G-Factor in Few-Layer Exfoliated MoS2 Devices. *J. Phys. Condens. Matter* **2025**, *37* (18), 185502. https://doi.org/10.1088/1361-648X/adc35d.
(20) Sharma, C. H.; Prada, M.; Schmidt, J.-H.; González Díaz-Palacio, I.; Stauber, T.; Taniguchi, T.; Watanabe, K.; Tiemann, L.; Blick, R. H. Transport Spectroscopy Study of Minibands in MoS 2 Moiré Superlattices. *Phys. Rev. B* **2024**, *109* (19), 195106. https://doi.org/10.1103/PhysRevB.109.195106.
(21) Zhou, C.; Wang, X.; Raju, S.; Lin, Z.; Villaroman, D.; Huang, B.; Chan, H. L. W.; Chan, M.; Chai, Y. Low Voltage and High ON/OFF Ratio Field-Effect Transistors Based on CVD MoS2and Ultra High-k Gate Dielectric PZT. *Nanoscale* **2015**, *7* (19), 8695–8700. https://doi.org/10.1039/c5nr01072a.
(22) Sharma, C. H.; Thalakulam, M. Split-Gated Point-Contact for Electrostatic Confinement of Transport in MoS2/h-BN Hybrid Structures. *Sci. Rep.* **2017**, *7* (1), 735. https://doi.org/10.1038/s41598-017-00857-7.
(23) Healy, B. F. M.; Pain, S. L.; Walker, M.; Grant, N. E.; Murphy, J. D. Impact of Co-Reactants in Atomic Layer Deposition of High-κ Dielectrics on Monolayer Molybdenum Disulfide. *ACS Appl. Nano Mater.* **2025**, *8* (14), 7334–7346. https://doi.org/10.1021/acsanm.5c00901.

(24) Sotthewes, K.; van Bremen, R.; Dollekamp, E.; Boulogne, T.; Nowakowski, K.; Kas, D.; Zandvliet, H. J. W.; Bampoulis, P. Universal Fermi-Level Pinning in Transition-Metal Dichalcogenides. *J. Phys. Chem. C* **2019**, *123* (9), 5411–5420. https://doi.org/10.1021/acs.jpcc.8b10971.
(25) Zhou, L.-X. Quasi-Bonding-Induced Gap States in Metal/Two-Dimensional Semiconductor Junctions: Route for Schottky Barrier Height Reduction. *Phys. Rev. B* **2022**, *105* (22). https://doi.org/10.1103/PhysRevB.105.224105.
(26) Yang, H.; Kim, N. Y. Material-Inherent Noise Sources in Quantum Information Architecture. *Materials* **2023**, *16* (7), 2561. https://doi.org/10.3390/ma16072561.
(27) Steinacker, P.; Dumoulin Stuyck, N.; Lim, W. H.; Tanttu, T.; Feng, M.; Serrano, S.; Nickl, A.; Candido, M.; Cifuentes, J. D.; Vahapoglu, E.; Bartee, S. K.; Hudson, F. E.; Chan, K. W.; Kubicek, S.; Jussot, J.; Canvel, Y.; Beyne, S.; Shimura, Y.; Loo, R.; Godfrin, C.; Raes, B.; Baudot, S.; Wan, D.; Laucht, A.; Yang, C. H.; Saraiva, A.; Escott, C. C.; De Greve, K.; Dzurak, A. S. Industry-Compatible Silicon Spin-Qubit Unit Cells Exceeding 99% Fidelity. *Nature* **2025**, *646* (8083), 81–87. https://doi.org/10.1038/s41586-025-09531-9.
(28) Burkard, G.; Ladd, T. D.; Pan, A.; Nichol, J. M.; Petta, J. R. Semiconductor Spin Qubits. *Rev. Mod. Phys.* **2023**, *95* (2), 025003. https://doi.org/10.1103/RevModPhys.95.025003.
(29) Van Der Wiel, W. G.; De Franceschi, S.; Elzerman, J. M.; Fujisawa, T.; Tarucha, S.; Kouwenhoven, L. P. Electron Transport through Double Quantum Dots. *Rev. Mod. Phys.* **2002**, *75* (1), 1–22. https://doi.org/10.1103/RevModPhys.75.1.
(30) Kumbhakar, P.; Shanmugam, A.; Sharma, C. H.; Reno, J. L.; Thalakulam, M. Quantum Point Contact Galvanically Coupled to Planar Superconducting Resonator: A Shot-Noise-Limited Broad-Band Electrical Amplifier. *Quantum Sci. Technol.* **2021**, *6* (4), 045006. https://doi.org/10.1088/2058-9565/ac107f.
(31) Shanmugam, A.; Kumbhakar, P.; Sundaresan, H.; Sunny, A. A.; Reno, J. L.; Thalakulam, M. GHz Operation of a Quantum Point Contact Using Stub-Impedance Matching Circuit. *Phys. Open* **2023**, *17*, 100181. https://doi.org/10.1016/j.physo.2023.100181.
(32) Sakudo, T.; Unoki, H. Dielectric Properties of SrTiO3 at Low Temperatures. *Phys. Rev. Lett.* **1971**, *26* (14), 851–853. https://doi.org/10.1103/PhysRevLett.26.851.
(33) Müller, K. A.; Burkard, H. SrTi O 3 : An Intrinsic Quantum Paraelectric below 4 K. *Phys. Rev. B* **1979**, *19* (7), 3593–3602. https://doi.org/10.1103/PhysRevB.19.3593.
(34) Muragesh, P.; Sundaresan, H.; Thalakulam, M. Cryogenic Microwave Frequency Combs Based on Quantum Paraelectric Superconducting Resonators. arXiv May 13, 2026. https://doi.org/10.48550/arXiv.2605.13571.
(35) Sundaresan, H.; Muragesh, P.; Thalakulam, M. Cryogenic Microwave Mixing and Phase Control in Superconducting Quantum Paraelectric Resonator. *Cell Rep. Phys. Sci.* **2026**, *7* (7). https://doi.org/10.1016/j.xcrp.2026.103387.
(36) Novoselov, K. S.; Jiang, D.; Schedin, F.; Booth, T. J.; Khotkevich, V. V.; Morozov, S. V.; Geim, A. K. Two-Dimensional Atomic Crystals. *Proc. Natl. Acad. Sci.* **2005**, *102* (30), 10451–10453. https://doi.org/10.1073/pnas.0502848102.
(37) Castellanos-Gomez, A.; Buscema, M.; Molenaar, R.; Singh, V.; Janssen, L.; Van Der Zant, H. S. J.; Steele, G. A. Deterministic Transfer of Two-Dimensional Materials by All-Dry Viscoelastic Stamping. *2D Mater.* **2014**, *1* (1), 011002. https://doi.org/10.1088/2053-1583/1/1/011002.
(38) Jana, S. P.; Shivangi; Gupta, S.; Gupta, A. K. Enhanced Performance of MoS$_2$/SiO$_2$ Field-Effect Transistors by Hexamethyldisilazane (HMDS) Encapsulation. arXiv March 9, 2024. https://doi.org/10.48550/arXiv.2403.05885.
(39) Liu, N.; Baek, J.; Kim, S. M.; Hong, S.; Hong, Y. K.; Kim, Y. S.; Kim, H.-S.; Kim, S.; Park, J. Improving the Stability of High-Performance Multilayer $MoS_2$ Field-Effect Transistors. *ACS Appl. Mater. Interfaces* **2017**, *9* (49), 42943–42950. https://doi.org/10.1021/acsami.7b16670.
(40) Suleman, M.; Lee, S.; Kim, M.; Nguyen, V. H.; Riaz, M.; Nasir, N.; Kumar, S.; Park, H. M.; Jung, J.; Seo, Y. NaCl-Assisted Temperature-Dependent Controllable Growth of Large-Area MoS2 Crystals Using Confined-Space CVD. *ACS Omega* **2022**, *7* (34), 30074–30086. https://doi.org/10.1021/acsomega.2c03108.
(41) Singh, A.; Singh, A. K. Origin of $n$-Type Conductivity of Monolayer ${\mathrm{MoS}}_{2}$. *Phys. Rev. B* **2019**, *99* (12), 121201. https://doi.org/10.1103/PhysRevB.99.121201.
(42) Park, Y.; Li, N.; Jung, D.; Singh, L. T.; Baik, J.; Lee, E.; Oh, D.; Kim, Y. D.; Lee, J. Y.; Woo, J.; Park, S.; Kim, H.; Lee, G.; Lee, G.; Hwang, C.-C. Unveiling the Origin of N-Type Doping of Natural MoS2: Carbon. *Npj 2D Mater. Appl.* **2023**, *7* (1), 60. https://doi.org/10.1038/s41699-023-00424-x.
(43) Salahuddin, S.; Datta, S. Use of Negative Capacitance to Provide Voltage Amplification for Low Power Nanoscale Devices. *Nano Lett.* **2008**, *8* (2), 405–410. https://doi.org/10.1021/nl071804g.
(44) Ayari, A.; Cobas, E.; Ogundadegbe, O.; Fuhrer, M. S. Realization and Electrical Characterization of Ultrathin Crystals of Layered Transition-Metal Dichalcogenides. *J. Appl. Phys.* **2007**, *101* (1), 014507. https://doi.org/10.1063/1.2407388.
(45) Huang, X.; Yao, Y.; Peng, S.; Zhang, D.; Shi, J.; Jin, Z. Effects of Charge Trapping at the MoS2–SiO2 Interface on the Stability of Subthreshold Swing of MoS2 Field Effect Transistors. *Materials* **2020**, *13* (13), 2896. https://doi.org/10.3390/ma13132896.
(46) Sun, Y.; Jiang, L.; Wang, Z.; Hou, Z.; Dai, L.; Wang, Y.; Zhao, J.; Xie, Y.-H.; Zhao, L.; Jiang, Z.; Ren, W.; Niu, G. Multiwavelength High-Detectivity MoS2 Photodetectors with Schottky Contacts. *ACS Nano* **2022**, *16* (12), 20272–20280. https://doi.org/10.1021/acsnano.2c06062.
(47) Ortiz-Conde, A.; García Sánchez, F. J.; Liou, J. J.; Cerdeira, A.; Estrada, M.; Yue, Y. A Review of Recent MOSFET Threshold Voltage Extraction Methods. *Microelectron. Reliab.* **2002**, *42* (4–5), 583–596. https://doi.org/10.1016/S0026-2714(02)00027-6.
(48) Leong, W. S.; Li, Y.; Luo, X.; Nai, C. T.; Quek, S. Y.; Thong, J. T. L. Tuning the Threshold Voltage of MoS2 Field-Effect Transistors via Surface Treatment. *Nanoscale* **2015**, *7* (24), 10823–10831. https://doi.org/10.1039/C5NR00253B.
(49) Radisavljevic, B.; Radenovic, A.; Brivio, J.; Giacometti, V.; Kis, A. Single-Layer MoS2 Transistors. *Nat. Nanotechnol.* **2011**, *6* (3), 147–150. https://doi.org/10.1038/nnano.2010.279.
(50) Chen, Y.; Wang, X.; Wang, P.; Huang, H.; Wu, G.; Tian, B.; Hong, Z.; Wang, Y.; Sun, S.; Shen, H.; Wang, J.; Hu, W.; Sun, J.; Meng, X.; Chu, J. Optoelectronic Properties of Few-Layer $MoS_2$ FET Gated by Ferroelectric Relaxor Polymer. *ACS Appl. Mater. Interfaces* **2016**, *8* (47), 32083–32088. https://doi.org/10.1021/acsami.6b10206.
(51) Gevorgian, S.; Berg, H.; Jacobsson, H.; Lewin, T. Application Notes - Basic Parameters of Coplanar-Strip Waveguides on Multilayer Dielectric/Semiconductor Substrates, Part 1: High Permittivity Superstrates. *IEEE Microw. Mag.* **2003**, *4* (2), 60–70. https://doi.org/10.1109/MMW.2003.1201599.
(52) Fassler, A.; Majidi, C. Soft-Matter Capacitors and Inductors for Hyperelastic Strain Sensing and Stretchable Electronics. *Smart Mater. Struct.* **2013**, *22* (5), 055023. https://doi.org/10.1088/0964-1726/22/5/055023.
(53) Novoselov, K. S.; Jiang, D.; Schedin, F.; Booth, T. J.; Khotkevich, V. V.; Morozov, S. V.; Geim, A. K. Two-Dimensional Atomic Crystals.
(54) Qiu, H.; Pan, L.; Yao, Z.; Li, J.; Shi, Y.; Wang, X. Electrical Characterization of Back-Gated Bi-Layer MoS2 Field-Effect Transistors and the Effect of Ambient on Their Performances. *Appl. Phys. Lett.* **2012**, *100* (12), 123104. https://doi.org/10.1063/1.3696045.
(55) Late, D. J.; Liu, B.; Matte, H. S. S. R.; Dravid, V. P.; Rao, C. N. R. Hysteresis in Single-Layer $MoS_2$ Field Effect Transistors. *ACS Nano* **2012**, *6* (6), 5635–5641. https://doi.org/10.1021/nn301572c.
(56) Park, M. J.; Yi, S.-G.; Kim, J. H.; Yoo, K.-H. Metal–Insulator Crossover in Multilayered $MoS_2$. *Nanoscale* **2015**, *7* (37), 15127–15133. https://doi.org/10.1039/C5NR05223H.
(57) Radisavljevic, B.; Kis, A. Mobility Engineering and a Metal–Insulator Transition in Monolayer MoS2. *Nat. Mater.* **2013**, *12* (9), 815–820. https://doi.org/10.1038/nmat3687.

(58) Müller, K. A.; Burkard, H. SrTi O 3 : An Intrinsic Quantum Paraelectric below 4 K. *Phys. Rev. B* **1979**, *19* (7), 3593–3602. https://doi.org/10.1103/PhysRevB.19.3593.
(59) Sakudo, T.; Unoki, H. Dielectric Properties of SrTiO3 at Low Temperatures. *Phys. Rev. Lett.* **1971**, *26* (14), 851–853. https://doi.org/10.1103/PhysRevLett.26.851.
(60) Rytz, D.; Höchli, U. T.; Bilz, H. Dielectric Susceptibility in Quantum Ferroelectrics. *Phys. Rev. B* **1980**, *22* (1), 359–364. https://doi.org/10.1103/PhysRevB.22.359.
(61) McGuire, F. A.; Cheng, Z.; Price, K.; Franklin, A. D. Sub-60 mV/Decade Switching in 2D Negative Capacitance Field-Effect Transistors with Integrated Ferroelectric Polymer. *Appl. Phys. Lett.* **2016**, *109* (9), 093101. https://doi.org/10.1063/1.4961108.
(62) Yang, A. J.; Han, K.; Huang, K.; Ye, C.; Wen, W.; Zhu, R.; Zhu, R.; Xu, J.; Yu, T.; Gao, P.; Xiong, Q.; Renshaw Wang, X. Van Der Waals Integration of High-κ Perovskite Oxides and Two-Dimensional Semiconductors. *Nat. Electron.* **2022**, *5* (4), 233–240. https://doi.org/10.1038/s41928-022-00753-7.
(63) Liu, X.; He, J.; Liu, Q.; Tang, D.; Wen, J.; Liu, W.; Yu, W.; Wu, J.; He, Z.; Lu, Y.; Zhu, D.; Liu, W.; Cao, P.; Han, S.; Ang, K.-W. Low Temperature Carrier Transport Study of Monolayer MoS2 Field Effect Transistors Prepared by Chemical Vapor Deposition under an Atmospheric Pressure. *J. Appl. Phys.* **2015**, *118* (12), 124506. https://doi.org/10.1063/1.4931617.
(64) Fivaz, R.; Mooser, E. Mobility of Charge Carriers in Semiconducting Layer Structures. *Phys. Rev.* **1967**, *163* (3), 743–755. https://doi.org/10.1103/PhysRev.163.743.
(65) Kaasbjerg, K.; Thygesen, K. S.; Jacobsen, K. W. Phonon-Limited Mobility in n -Type Single-Layer MoS 2 from First Principles. *Phys. Rev. B* **2012**, *85* (11), 115317. https://doi.org/10.1103/PhysRevB.85.115317.
(66) Jariwala, D.; Sangwan, V. K.; Late, D. J.; Johns, J. E.; Dravid, V. P.; Marks, T. J.; Lauhon, L. J.; Hersam, M. C. Band-like Transport in High Mobility Unencapsulated Single-Layer MoS2 Transistors. *Appl. Phys. Lett.* **2013**, *102* (17), 173107. https://doi.org/10.1063/1.4803920.
(67) Gurusinghe, M. N.; Davidsson, S. K.; Andersson, T. G. Two-Dimensional Electron Mobility Limitation Mechanisms in Al x Ga 1 − x N ⁄ GaN Heterostructures. *Phys. Rev. B* **2005**, *72* (4), 045316. https://doi.org/10.1103/PhysRevB.72.045316.
(68) Shanmugam, A.; Thekke Purayil, M. A.; Dhurjati, S. A.; Thalakulam, M. Physical Vapor Deposition-Free Scalable High-Efficiency Electrical Contacts to $MoS_2$. *Nanotechnology* **2024**, *35* (11), 115201. https://doi.org/10.1088/1361-6528/ad12e4.
(69) Qiu, H.; Xu, T.; Wang, Z.; Ren, W.; Nan, H.; Ni, Z.; Chen, Q.; Yuan, S.; Miao, F.; Song, F.; Long, G.; Shi, Y.; Sun, L.; Wang, J.; Wang, X. Hopping Transport through Defect-Induced Localized States in Molybdenum Disulphide. *Nat. Commun.* **2013**, *4* (1), 2642. https://doi.org/10.1038/ncomms3642.

Supporting Information for:

# Coplanar Lateral Gating $MoS_2$ on $SrTiO_3$: A Unified Platform for Classical and Quantum Devices

*Prasad Muragesh[#], Manav Murali[#], Venkatesha Modur Ramachandra, and Madhu Thalakulam[*]*

Indian Institute of Science Education and Research Thiruvananthapuram, Thiruvananthapuram, Kerala 695551, India

[#]*equally contributed to the work*

[*]*madhu@iisertvm.ac.in*

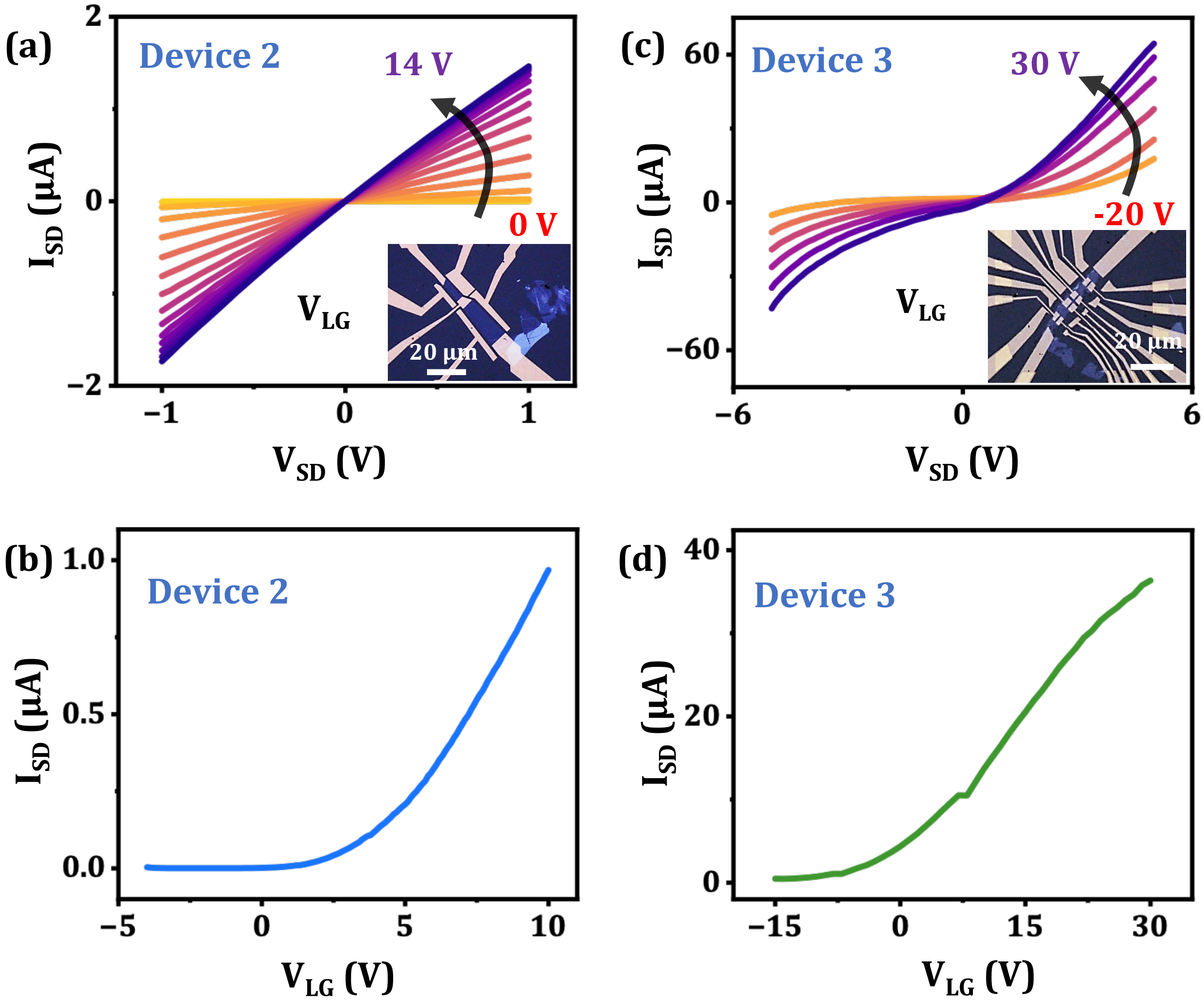

**SI-1: Room Temperature Transport Data for other Laterally Gated $MoS_2$/STO Devices (Device 2, Device 3) :** (a) I-V characteristics of Device 2, recorded for the LG configuration, where $V_{LG} = 0\,V\ to\ 14\,V$ while keeping the $V_{BG} = 0\,V$, ***Inset:*** optical microscope image of the device; (b) Transfer curves recorded for device 2 with $V_{SD} = 1\,V$ and $V_{LG} = -4\,V$ to $10\,V$ (c) I-V characteristics of device 3, $V_{LG} = -20\,V$ to $30\,V$, **Inset:** optical microscope image of the device. (d) Transfer curves recorded for device 3 with $V_{SD} = 2\,V$ and $V_{LG} = -15\,V$ to $30\,V$.

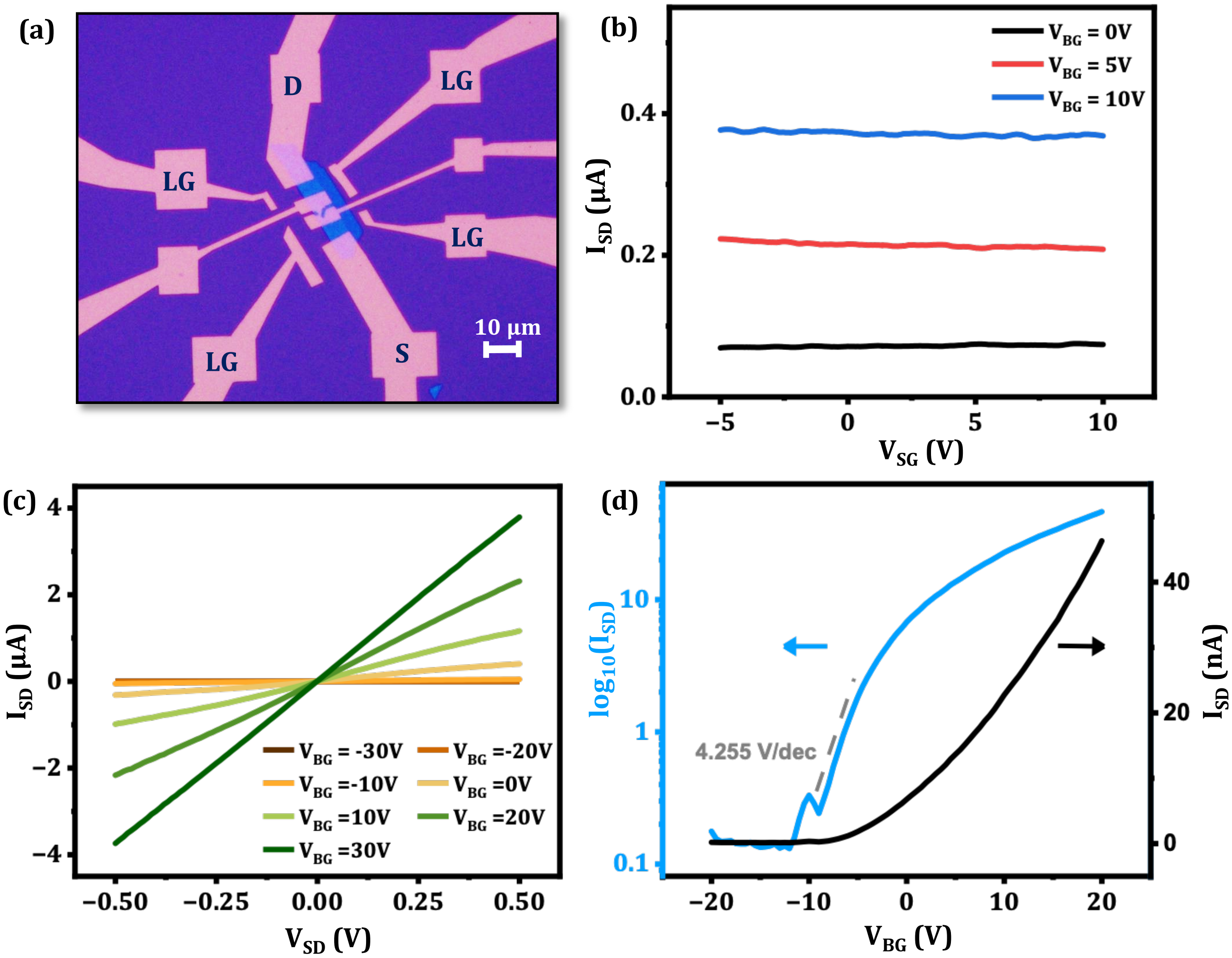

**SI-2: Room-temperature study of the lateral-gate architecture of $MoS_2$ on a $SiO_2$ substrate.** (a) Optical image of the fabricated device on the $SiO_2$ substrate; (b) $I_{SD}$-$V_{SG}$ sweeps measured for different back-gate voltages $V_{BG}$ at $V_{SD} = 10\ mV$. The side-gate architecture was found to be ineffective in the $MoS_2/SiO_2$ device configuration, confirming the limited efficiency of the architecture in a low dielectric-constant environment. (c) The BG-dependent I-V characteristics measured for $V_{BG}$ varying from $-30\ V$ to $30\ V$ while keeping $V_{LG} = 0V$. (d) $I_{SD}$-$V_{BG}$ transfer characteristics measured at $V_{SD} = 10\ mV$, with the extracted subthreshold swing estimated to be $\sim 4.255\ V/dec$.

**SI-3: Details on electrostatic Simulations:**

The simulation was built using FEniCSx [S1]. The entire computational domain was discretized into structured hexahedral elements, roughly 12 millions of them, and the resulting linear systems were solved at each step using PETSc [S2] with an algebraic multigrid preconditioner. The device we simulated has three regions stacked vertically. At the bottom sits a 100 $\mu m$ thick STO substrate. On top of it, right at the surface (z = 0), lies a 42 $nm$ thick multilayer $MoS_2$ channel, and above everything is a 3 $\mu m$ vacuum region. The in-plane layout of the device - the exact shape of the $MoS_2$ flake, the positions of the gate electrodes, source and drain contacts was taken directly from the GDSII file used for device fabrication. We studied two gating configurations: one with coplanar surface gates that sit on the same plane as the channel, separated from it by approximately 1 $\mu m$ air gaps, and another with a global back gate at the bottom of the STO substrate, 500 $\mu m$ away from the channel.

The physics of the simulation is governed by the nonlinear Poisson equation. At every point in the 3D domain, the electrostatic potential V must satisfy

$$\nabla \cdot [\varepsilon_0\, \hat{\varepsilon}_{\mathrm{r}}(r)\, \nabla V] \;=\; -\rho(V,T) \tag{1}$$

The permittivity $\hat{\varepsilon}_r$ is 1 in vacuum, anisotropic in $MoS_2$ ($\varepsilon_{\parallel} = 15.9$, $\varepsilon_{\perp} = 6.9$ [S3]), and isotropic but strongly temperature-dependent in STO. The STO dielectric constant is modelled using the Barrett formula [S9]:

$$\varepsilon(T) \;=\; \frac{C}{\frac{T_1}{2}\coth\left(\frac{T_1}{2T}\right) - T_0} \tag{2}$$

with $C = 8.05 \times 10^4$ K, $T_1 = 84$ K, and $T_0 = 38.6$ K, yielding $\varepsilon \approx 306$ at 300 K, ~1754 at 77 K, and ~23,000 at 4 K. This enormous, temperature-tunable permittivity causes electric fields to route preferentially through the substrate rather than vacuum.

**Charge model.** The charge density ρ is nonzero only in the $MoS_2$ channel:

$$\rho \;=\; e[N_{\mathrm{D}}^{+}(V,T) \;-\; n(V,T)] \tag{3}$$

$MoS_2$ is natively n-type due to Sulphur vacancies ($N_D = 1.2 \times 10^{13} cm^{-2}$ [S7]) with deep donor levels at $E_D \;=\; 0.46\; eV$ below the conduction band [S8]. The free electron density uses full Fermi–Dirac statistics:

$$n(V,T) \;=\; N_{\mathrm{C}}(T) \;\cdot\; \mathcal{F}_{1/2}\left(\frac{eV}{k_{\mathrm{B}}T}\right) \tag{4}$$

where $N_C = 2(m_{\mathrm{dos}}\, k_B T \,/\, 2\pi\hbar^2)^{3/2}$ is the effective density of states with $m_{\mathrm{dos}} = g_v^{2/3} \cdot m^*$, using $m^* = 0.48\, m_0$ [S4] and valley degeneracy $g_v = 6$ [S5] appropriate for bulk-like 42 $nm$ $MoS_2$. The Fermi–Dirac integral $\mathcal{F}_{1/2}$ is evaluated using the Bednarczyk–Bednarczyk rational approximation [S6]. The ionized donor density follows standard deep-donor occupation statistics:

$$N_{\mathrm{D}}^{+} \;=\; \frac{N_{\mathrm{D}}}{1 + g_{\mathrm{D}}\exp\left(\frac{eV + E_{\mathrm{D}}}{k_{\mathrm{B}}T}\right)} \tag{5}$$

with spin degeneracy $g_D = 2$. Since ρ depends on $V$ and vice versa, the problem is solved self-consistently.

**Boundary conditions and solver.** Gate electrodes are fixed at $V_g$ (Dirichlet); source/drain contacts at applied voltage plus a built-in potential $V_{bi}$ determined from charge neutrality $n(V_{bi}) = N_D^+(V_{bi})$ at each temperature; all other boundaries use Neumann conditions ($\partial V / \partial n = 0$). The nonlinear system is solved by Newton–Raphson iteration, starting from a Laplace-equation initial guess (ρ = 0). The mesh uses 0.7 nm vertical spacing in the channel, smoothly coarsening into the substrate and vacuum, with ~150 nm in-plane spacing.

**Simulation sweep.** We performed 507 simulations for the coplanar gate configuration (13 temperatures × 13 gate voltages × 3 source-drain biases) and 48 simulations for the back gate geometry (4 temperatures × 4 gate voltages × 3 source-drain biases), extracting the full 3D potential $V(\boldsymbol{r})$ at each parameter point.

**Material parameters used in the simulation.**

| Parameter | Symbol | Value | Ref. |
|---|---|---|---|
| **Band gap (indirect)** | $E^g$ | 1.29 eV | [S5] |
| **Effective mass** | m* | 0.48 $m_0$ | [S4] |
| **Valley degeneracy** | $g_v$ | 6 | [S5] |
| **In-plane permittivity** | ε‖ | 15.9 | [S3] |
| **Out-of-plane permittivity** | ε⊥ | 6.9 | [S3] |
| **Channel thickness** | t | 42 nm | Expt. |
| **2D donor density** | $N^D$ | $1.2 \times 10^{13}$ cm$^{-2}$ | [S7] |
| **Donor ionization energy** | $E^D$ | 0.46 eV | [S8] |
| **Donor degeneracy** | $g^D$ | 2 | — |
| **Barrett C** | C | $8.05 \times 10^4$ K | [S9] |
| **Barrett $T_1$** | $T_1$ | 84 K | [S9] |
| **Barrett $T_0$** | $T_0$ | 38.6 K | [S9] |

## References:

[S1] M. S. Alnes et al., "The FEniCS Project Version 1.5," Archive of Numerical Software 3, 9 (2015).
[S2] S. Balay et al., "PETSc Users Manual," ANL-95/11, Revision 3.22, Argonne National Laboratory (2024).
[S3] A. Laturia, M. L. Van de Put, and W. G. Vandenberghe, npj 2D Mater. Appl. 2, 6 (2018).
[S4] K. Kaasbjerg, K. S. Thygesen, and K.-W. Jacobsen, Phys. Rev. B 85, 115317 (2012).
[S5] T. Cheiwchanchamnangij and W. R. L. Lambrecht, Phys. Rev. B 85, 205302 (2012).
[S6] D. Bednarczyk and J. Bednarczyk, Phys. Lett. A 64, 409 (1978).
[S7] J. Hong et al., Nat. Commun. 6, 6293 (2015).
[S8] H. Qiu et al., Nat. Commun. 4, 2642 (2013).
[S9]K. A. Müller and H. Burkard, Phys. Rev. B 19, 3593 (1979).

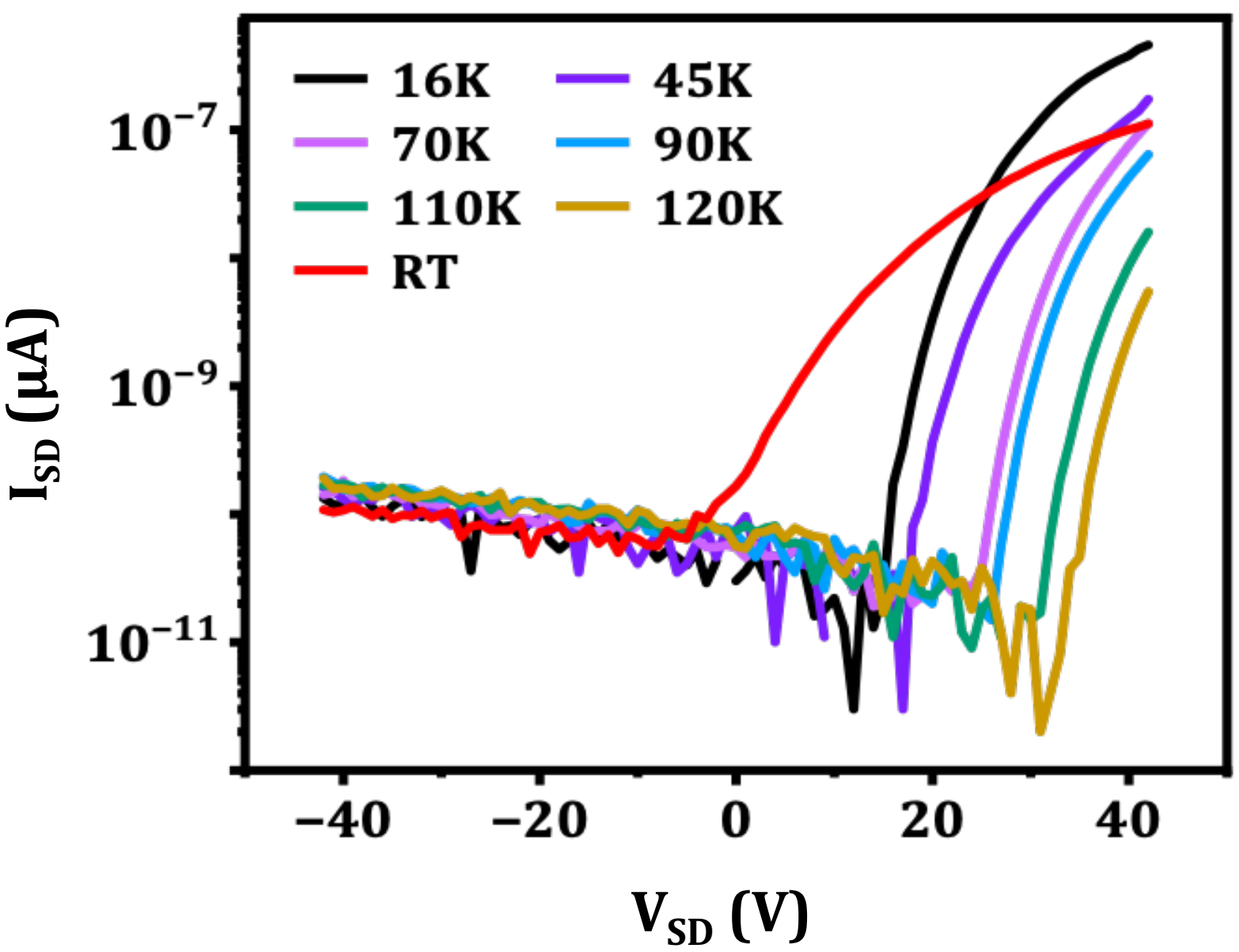

**SI-4: Temperature dependent transfer characteristic of MoS2/STO back gate.**

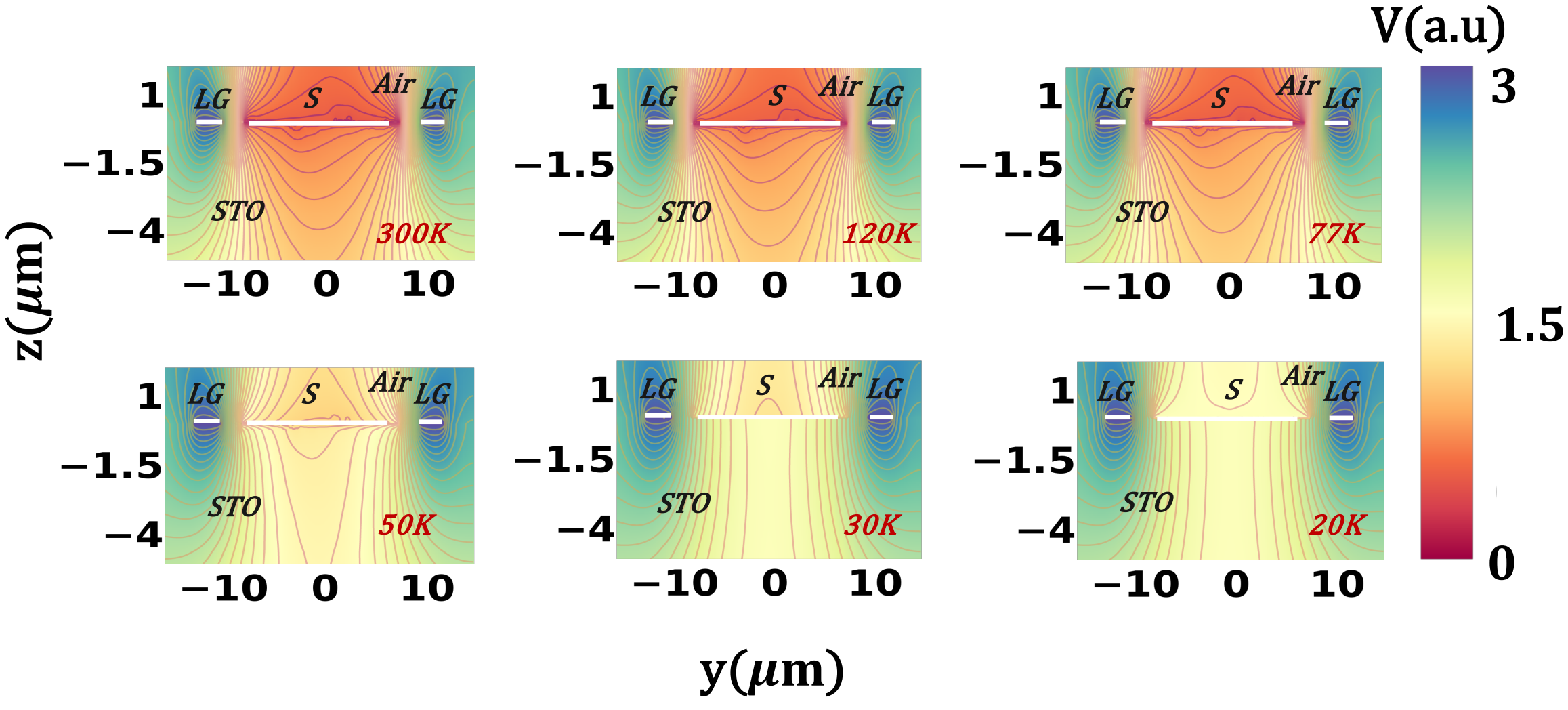

**SI-5: Simulation electrostatic potential for additional temperatures.**